\documentclass[aip,graphicx]{revtex4-2}

\usepackage{amsmath}
\usepackage{xcolor}
\usepackage{soul}
\newcommand{\etal}{ \textit{et~al. }}

\begin{document}

\title{To be published in Applied Physics Letters  \\
\vspace{10mm}
Comment on “Color astrophotography with a 100 mm-diameter f/2 polymer flat lens” Appl. Phys. Lett. 126, 051701 (2025) }
\author{G. K. Skinner}
 \email{g.k.skinner@bham.ac.uk.}
\affiliation{ Honorary Senior Research Fellow, University of Birmingham, UK.}%
\author{J. F. Krizmanic}
\affiliation{Senior Astrophysicist, Code 661, NASA GSFC, USA}

\date{\today}

\maketitle 

Majumder \etal\cite{2025ApPhL.126e1701M} advocate  the use for astronomy of large, thin, diffractive   lenses that are optimised at a number of wavelengths over a prescribed bandwidth. In an earlier paper by the same group \cite{BanerjiSWIR}  the efficiency of such lenses was over-stated, as discussed in Ref  \onlinecite{Skinner:24}. The efficiencies quoted in this latest letter may or may not be correct, but they are defined and measured in such a way that they are not useful measures of how much of the incident radiation is brought to a focus. The actual focussing efficiency is much lower than the $\sim$ 3\% quoted in the letter. 
A reading of the letter without a detailed and knowledgeable scrutiny of the 50+ pages of supplementary information could lead to an impression that such lenses have much more potential for astronomy than is likely to be the case. There are indications in the form of the publicity that has been given to the letter ({\it{e.g.}} \onlinecite{spacedotcom}) that this has already happened.

The {\it focussing efficiency} of a lens may be defined as the fraction of the incident radiation  on the lens that is focused  inside a circle of a specified  diameter, $d$ in the image plane. Typically  $d$ is related  to the size of the focal spot, say 3 times  the FWHM (Full Width at Half Maximum). For brevity, comments here are restricted to the information presented on lens MDL-100A for which, based on Figs. S3.7 and S4.11, this means inside $d \sim$ 5 to 50 microns depending on whether the FWHM considered is  that observed or that of an  ideal one (Airy disc).

 Instead the authors choose to quote the {\it integrated efficiency}, $\eta_{int}$,  defined by  Buralli and Morris
\cite{1992ApOpt314389B} 
in the different context of quantifying degradation of a diffraction grating OTF by spurious scattering. In  S11 this is called the  “integrated focussing efficiency”  
 but it has little to do with focussing. It refers to the flux in a large patch in the image plane, not in a focal spot of a few microns. The simulations in Fig S11.2 used a region 10 mm in diameter while experimental results measured  the energy falling on a detector of area 9.7 x 9.7 mm (S120VC datasheet).

An attempt to reconstruct an enclosed energy plot from simulated data  
 in Fig. S7.1 suggests that less than 1\% of the  energy within 5mm is actually in the focussed peak. Measurements  in Fig. S4.4 indicate an even lower fraction.
Thus the, already low, quoted   “integrated efficiency” of  3\% corresponds to a true focussing efficiency of $<0.03\%$. Worse, this figure is the average over 8 specific wavelengths which were among those for which the lens was optimised. Experience with designs of similar thin lenses suggests that the focussing efficiency at intermediate wavelengths will be far lower than at those wavelengths.  
Had similar large area measurements been made for the comparison refractive lens then  $\eta_{int}$ would  have been found to be close to 100\%  despite chromatic aberration spreading the focus to several mm diameter.

Low efficiency is not unexpected. Apart from the intrinsic limitations of thin diffractive lenses optimised at a set of discrete wavelengths\cite{Skinner:24}, another reason why the focussing efficiency is extremely low is that, as acknowledged by the authors, the  $5\mu m$ rings are far too wide for efficient diffraction. 
With such rings the radial Nyquist frequency is $0.1\; \mu m^{-1}$, whereas for first order diffraction of $400-800\; nm$ radiation  from the edge of the lens to the focal point a frequency up to 6 time higher would be needed.

The reason that the lenses  may appear at first sight to offer promising performance is that any set of concentric rings that introduce even random phase shifts is likely to have a response described by a PSF with a central core typically of width comparable to the diffraction limit, though containing only a tiny fraction of the flux. The great majority of the energy lies in extended wings to the PSF, forming background fog and drastically reducing the image contrast. Diagrams in which colors indicate intensity may still seem impressive, depending on the normalisation and color mapping. 

Apart from good light-collecting power, another reason that large diameter optics are generally favoured for astronomy is because they potentially offer high angular resolution. In principle a lens as large as 100 mm could be capable of imaging with FWHM $\sim 1.5$ arcsec, corresponding to $\sim 1.5 \mu m$ at 200mm. The performance reported in Fig. S3.7 is  ‘only’ a factor of 2--3 times 
that. But this is true only for a very narrow band around each wavelength chosen for optimisation. The lines joining the points in that figure are misleading and Fig, S4.11 shows that at most wavelengths the FWHM  is worse than  the diffraction limit by a factor $\sim10$.  If an alternative measure of resolution,  Half-Power-Diameter, were used that factor would be many hundreds. Another way of measuring focussing quality is the Strehl ratio but the data reported in S1.2c must be erroneous as they are incompatible with the focussing efficiency.

Although the lens described does demonstrate focusing over a wide wavelength band, the focus is imperfect and  the amount of light focussed is a minute fraction of that incident, the remainder being spread as fog over the focal plane. With poor contrast, with only the resolution of a 1 cm lens,  and with a focussed energy equal to that of a mm scale one, practical applications in astronomy of lenses such as that described seem unlikely.




\bibliography{flat_lenses_plus_a}

\end{document}